\documentclass[runningheads,a4paper]{llncs}

\usepackage{amssymb}
\usepackage{graphicx}
\usepackage{epstopdf}
\usepackage{url}
\urldef{\mailsa}\path|{alfred.hofmann, ursula.barth, ingrid.haas, frank.holzwarth,|
\urldef{\mailsb}\path|anna.kramer, leonie.kunz, christine.reiss, nicole.sator,|
\urldef{\mailsc}\path|erika.siebert-cole, peter.strasser, lncs}@springer.com|
\newcommand{\keywords}[1]{\par\addvspace\baselineskip
\noindent\keywordname\enspace\ignorespaces#1}

\begin{document}

\mainmatter  % start of an individual contribution

\title{Towards the Formal Reliability Analysis of Oil and Gas Pipelines}

\titlerunning{Lecture Notes in Computer Science: Authors' Instructions}


\author{Waqar Ahmed$^1$ %
%\thanks{Please note that the LNCS Editorial assumes that all authors have used
%the western naming convention, with given names preceding surnames. This determines
%the structure of the names in the running heads and the author index.}%
\and Osman Hasan$^1$ \and Sofiene Tahar$^2$ \and\\
Mohammad Salah Hamdi$^3$}

\institute{$^1$School of Electrical Engineering and Computer Science (SEECS)\\
National University of Sciences and Technology (NUST)\\
Islamabad, Pakistan \\
\email{ \{12phdwahmad,osman.hasan\}@seecs.nust.edu.pk  }\\
$^2$Electrical and Computer Engineering Department\\
Concordia University, Montreal, Canada \\
\email{tahar@ece.concordia.ca}\\
$^3$Information Systems Department \\
Ahmed Bin Mohammed Military College, Doha,  Qatar \\
\email{mshamdi@abmmc.edu.qa}
}


%\toctitle{Lecture Notes in Computer Science}
%\tocauthor{Authors' Instructions}
\maketitle

\begin{abstract}
It is customary to assess the reliability of underground oil and gas pipelines in the presence of excessive loading and corrosion effects to ensure a leak-free transport of hazardous materials. The main idea behind this reliability analysis is to model the given pipeline system as a Reliability Block Diagram (RBD) of segments such that the reliability of an individual pipeline segment can be represented by a random variable. Traditionally, computer simulation is used to perform this reliability analysis but it provides approximate results and requires an enormous amount of CPU time for attaining reasonable estimates. Due to its approximate nature, simulation is not very suitable for analyzing safety-critical systems like oil and gas pipelines, where even minor analysis flaws may result in catastrophic consequences. As an accurate alternative, we propose to use a higher-order-logic theorem prover (HOL) for the reliability analysis of pipelines. As a first step towards this idea, this paper provides a higher-order-logic formalization of reliability and the series RBD using the HOL theorem prover. For illustration, we present the formal analysis of a simple pipeline that can be modeled as a series RBD of segments with exponentially distributed failure times.
\keywords{Reliability Block Diagrams, Formal Methods, Theorem Proving, Oil and Gas pipeline}
\end{abstract}

\section{Introduction}

On April 20, 2010, methane gas leakage on the Deepwater Horizon oil rig operated by Transocean, a subcontractor of British Petroleum (BP), caused a big explosion \cite{bps_13b}. This leakage not only killed 11 workers instantly but destroyed and sank the rig, and caused millions of gallons of oil to pour into the Gulf of Mexico. The gushing well, about a mile under the sea, was finally brought under control after more than three months of frenetic attempts. The spill, which is considered to be the largest accidental marine oil spill in the history of the petroleum industry, caused extensive damage to marine and wildlife habitats as well as the Gulf's fishing and tourism industries and its impact still continues. Just like the BP pipeline, there are tens of thousands of miles long oil and gas pipelines around the world.  All of these pipelines are aging and are becoming more and more susceptible to failures, which may lead to disasters like the BP one. Hence, it is very important to do rigorous reliability analysis of oil and gas pipelines to detect and rectify potential problems.

The reliability analysis of a pipeline system involves a three-step process: (i) partitioning the given pipeline into segments and constructing its equivalent reliability block diagram (RBD), (ii) assessing the reliability of the individual segments and (iii) evaluating the reliability of the complete pipeline system based on the RBD  and the reliability of its individual segments. The reliability of an individual segment is usually expressed in terms of its failure rate $\lambda$ and a random variable, like exponential \cite{Zhang_08} or Weibull random variable \cite{Kolowrocki_09}, which models the failure time. A single oil or gas pipeline can be simply modeled as a series RBD \cite{Zhang_08}. However, in many cases, these pipeline systems have either reserved components or subsystems and such pipeline systems exhibit a combination of series and parallel RBDs \cite{Soszynska_10}.

The reliability analysis of oil and gas pipelines has predominantly been accomplished by first gathering data
from in-line inspection tools to detect cracks, corrosion or damage \cite{pipe_integ_sol_13,pipe_check_13}. This information is then manipulated using the
paper-and-pencil based analytical analysis and computer simulations to deliver diagnostics and
insightful pipeline integrity reports (e.g. \cite{panday_98,Zhang_08,Soszynska_10}). However, due to the complex nature of large pipeline system analysis, paper-and-pencil proof methods are error prone and the exhaustive testing of all possible system behaviors using simulation is almost impossible. Thus, these traditional analysis techniques cannot guarantee accurate results, which is a severe limitation in the case of oil and gas pipelines as an uncaught system bug may endanger human and animal life or lead to a significant financial loss.

The inaccuracy limitations of traditional analysis techniques can be overcome by using formal methods \cite{boca_09}, which use computerized mathematical reasoning to precisely model the system's intended behavior and to provide irrefutable proof that a system satisfies its requirements. Both model checking and theorem proving have
been successfully used for the precise probabilistic analysis of a
broad range of systems (e.g. \cite{hasan_ispass_08,KNP_10a,elleuch_11,hasan_jal_11,fruth_11}). However, to the best of our knowledge, no formal analysis approach has been used for the reliability analysis of oil and gas pipelines so far. The foremost requirement for conducting the formal reliability analysis
of underground oil and gas pipelines is the ability to formalize RBDs recursively and continuous random variables.
Model checking is a state-based formal method technique. The inherent limitations of model checking is the state-space explosion problem and the inability to model complex datatypes such as trees, lists and recursive definitions \cite{kaufman2004some}.
% and cannot support modeling  verifying continuous models.
 On the other hand, higher-order logic \cite{cebrown_07} is a system of deduction with a
precise semantics and can be used to formally model any system that
can be described mathematically including recursive definitions,  random variables, RBDs, and continuous components. Similarly, interactive theorem provers are computer based formal
reasoning tools that allow us to verify higher-order-logic properties under user
guidance. Higher-order-logic theorem provers can be used to reason about recursive definitions using induction methods \cite{kapur1996lemma}. Thus, higher-order-logic theorem proving can be used to conduct the formal analysis of oil and gas pipelines.

A number of higher-order-logic formalizations of probability theory are available in higher-order logic
(e.g. \cite{hurd_02,mhamdi_11,holzl_11}). Hurd's formalization of probability theory \cite{hurd_02} has been utilized to
verify sampling algorithms of a number of commonly used discrete
\cite{hurd_02} and continuous random variables \cite{hasan_cade_07}
based on their probabilistic and statistical properties
\cite{hasan_icnaam_07,hasan_fm_09}. Moreover, this formalization has been used to conduct the reliability analysis of a number of applications, such as memory arrays \cite{hasan_tc_10}, soft errors \cite{abbasi_13b} and electronic components \cite{abbasi_13}. However, Hurd's formalization of probability theory only supports having the whole universe as the probability space. This feature limits its scope and thus this probability theory cannot be used to formalize more than a single continuous random variable. Whereas, in the case of reliability analysis of pipelines, multiple continuous random variables are required. The recent formalizations of probability theory by Mhamdi
\cite{mhamdi_11} and H\"{o}lzl~\cite{holzl_11} are based on extended
real numbers (including $\pm\infty$) and provide the formalization of
Lebesgue integral for reasoning about advanced statistical
properties. These theories also allow using any arbitrary probability space that is a subset of the universe and thus are more flexible than Hurd's formalization. However, to the best of our knowledge, these foundational theories have not been used to formalize neither reliability and RBDs nor continuous random variables so far.

In this paper, we use Mhamdi's formalization of probability theory \cite{mhamdi_11}, which is available in the HOL theorem prover \cite{norris_hol}, to formalize reliability and the commonly used series RBD, where its individual segments are modeled as random variables. Our formalization includes various formally verified properties of reliability and series RBD that facilitate formal reasoning about the reliability of some simple pipelines using a theorem prover. To analyze more realistic models of pipelines, it is required to formalize other RBDs, such as parallel, series-parallel and parallel-series \cite{Bilinton_1992}. In order to illustrate the utilization and effectiveness of the proposed idea, we utilize the above mentioned formalization to analyze a simple pipeline that can be modeled as a series RBD with an exponential failure time for individual segments.

\section{Preliminaries}
\label{sec_2}
In this section, we give a brief introduction to theorem proving in general and the HOL theorem
prover in particular. The intent is to introduce the main
ideas behind this technique to facilitate the understanding of the paper for the reliability analysis community. We also summarize Mhamdi's formalization of probability theory \cite{mhamdi_11}  in this section.

\subsection{Theorem Proving}

Theorem proving \cite{gordon_89} is a widely used formal
verification technique. The system that needs to be analysed is
mathematically modelled in an appropriate logic and the properties of
interest are verified using computer based formal tools. The use of
formal logics as a modelling medium makes theorem proving a very
flexible verification technique as it is possible to formally verify
any system that can be described mathematically. The core of theorem
provers usually consists of some well-known axioms and primitive
inference rules. Soundness is assured as every new theorem must be
created from these basic or already proved axioms and primitive inference rules.

The verification effort of a theorem in a theorem prover varies from
trivial to complex depending on the underlying logic
\cite{harrison_96a}. For instance, first-order logic
\cite{fitting_96} utilizes the propositional calculus and terms
(constants, function names and free variables) and is
semi-decidable. A number of sound and complete first-order logic
automated reasoners are available that enable completely automated
proofs. More expressive logics, such as higher-order logic
\cite{cebrown_07}, can be used to model a wider range of problems
than first-order logic, but theorem proving for these logics cannot
be fully automated and thus involves user interaction to guide the
proof tools. For reliability analysis of pipelines, we need to formalize
(mathematically model) random variables as functions and their distribution properties are verified by quantifying over random variable functions. Henceforth, first-order logic does not support such formalization and we need to use higher-order logic to formalize the foundations of reliability analysis of pipelines.

\subsection{HOL Theorem Prover}

HOL is an interactive theorem prover developed at the
University of Cambridge, UK, for conducting proofs in higher-order logic.
It utilizes the simple type theory of Church \cite{church_40} along
with Hindley-Milner polymorphism \cite{milner_77} to implement
higher-order logic. HOL has been successfully used as a verification
framework for both software and hardware as well as a platform for
the formalization of pure mathematics.

The HOL core consists of only 5 basic
axioms and 8 primitive inference rules, which are implemented as ML
functions. Soundness is assured as every new theorem must be verified by applying these basic axioms and primitive inference rules or any other previously verified theorems/inference rules.

We utilized the HOL theories of
Booleans, lists, sets, positive integers, \emph{real} numbers,
measure and probability in our work. In fact, one of the primary
motivations of selecting the HOL theorem prover for our work was to
benefit from these built-in mathematical theories.
Table \ref{hol_basics} provides the mathematical interpretations of
some frequently used HOL symbols and functions, which are inherited
from existing HOL theories, in this paper.

\begin{table}[!htb]
\begin{center}

\begin{tabular}{|c|c|c|} \hline
{\bfseries HOL Symbol} & {\bfseries Standard Symbol} & {\bfseries Meaning}  \\
\hline \hline
             $\mathtt{\wedge}$& $and$  & Logical $and$   \\ \hline
             $\mathtt{\vee}$ & $or$  &  Logical $or$  \\ \hline
             $\mathtt{\neg}$ & $not$  &  Logical $negation$  \\ \hline
             $\mathtt{::}$  &  $cons$ & Adds a new element to a list  \\ \hline
             $\mathtt{++}$  &  $append$ & Joins two lists together   \\ \hline
             $\mathtt{HD\ L}$  &  $head$ & Head element of list $L$ \\ \hline
             $\mathtt{TL\ L}$  &  $tail$ & Tail of list $L$\\ \hline
             $\mathtt{EL\ n\ L}$  &  $element$ & $n^{th}$ element of list L \\ \hline
             $\mathtt{MEM\ a\ L}$  &  $member$ & True if $a$ is a member of list $L$\\ \hline
%             $\mathtt{LENGTH\ L}$  &  $length$ & Length of list $L$\\ \hline
             %$\mathtt{(a,b)}$& a x b  & A pair of two elements    \\ \hline
%             $\mathtt{FS}$& fst (a, b) = a  & First component of a pair   \\ \hline
%             $\mathtt{snd}$& snd (a, b) = b  & Second component of a pair   \\ \hline
             $\mathtt{\lambda x.t}$& $\lambda x.t$  &  Function that maps $x$ to $t(x)$  \\ \hline
             %$\mathtt{\{x|P(x)\}}$& $\{\lambda x.P(x)\}$  & Set of all $x$ such that $P(x)$ \\ \hline
%             $\mathtt{num}$  &  $\{0,1,2,\ldots\}$ & Positive Integers data type   \\ \hline
%             $\mathtt{real}$ &  All Real numbers &  Real data type  \\ \hline
             $\mathtt{SUC\ n}$& $n + 1$  &  Successor of a $num$  \\ \hline
             %$\mathtt{ln\ x}$ & $log_{e}(x)$  & Natural logarithm function   \\ \hline
             %$\mathtt{exp\ x}$ & $e^x$  & Exponential function   \\ \hline
             %$\mathtt{sqrt\ x}$ & $\sqrt{x}$  & Square root function   \\ \hline
            % $\mathtt{abs\ x}$ & $|x|$  & Absolute function   \\ \hline
             $\mathtt{lim(\lambda n.f(n))}$ & $\mathop {\lim }\limits_{n \to \infty } f(n)$  &  Limit of a $real$ sequence $f$  \\
%             \hline
%             $\mathtt{convergent(\lambda n.f(n))}$ & $\exists x. \mathop {\lim }\limits_{n \to \infty } f(n) = x$  &  $f$ is convergent \\
%             \hline
%             $\mathtt{suminf(\lambda n.f(n))}$ & $\mathop {\lim }\limits_{k \to \infty } \sum^{k}_{n=0} f(n)$  &  Infinite summation of $f$  \\ \hline
%             $\mathtt{summable(\lambda n.f(n))}$ & $\exists x. \mathop {\lim }\limits_{k \to \infty } \sum^{k}_{n=0} f(n) = x$  &  Summation of $f$ is convergent  \\ \hline

%             SUC$\ n$& $n + 1$  &  Successor of a $num$  \\ \hline
%             $m\ **\ n $& $m^{n}$  & $num$ $m$ raised to $num$ exponent $n$   \\ \hline
%             $\mathtt{inv\ x}$& $x^{-1}$  &   Multiplicative inverse of a $real$ $x$ \\ \hline
%             $\lambda x.t$& $\lambda x.t$  &  Function that maps $x$ to $t(x)$  \\ \hline
%             $lim(\lambda n.f(n))$ & $\mathop {\lim }\limits_{n \to \infty } f(n)$  &  Limit of a $real$ sequence $f$  \\ \hline
%             $\{x|P(x)\}$& $\{\lambda x.P(x)\}$  & Set of all $x$ that satisfy the condition $P$ \\ \hline
%             $(a,b)$& a x b  & A mathematical pair of two elements    \\ \hline
%             @x.t & $\varepsilon x.t$ & An x such that t becomes
%             true
%\\ \hline
\hline
\end{tabular}
\caption{HOL Symbols and Functions}
 \label{hol_basics}
\end{center}
\end{table}

\subsection{Probability Theory and Random Variables in HOL}

Mathematically, a measure space is defined as a triple ($\Omega,\Sigma, \mu$), where
 $\Omega$ is a set, called the sample space, $\Sigma$ represents a $\sigma$-algebra of subsets of
$\Omega$, where the subsets are usually referred to as measurable sets, and $\mu$ is a measure with domain
$\Sigma$. A probability space is a measure space ($\Omega,\Sigma, Pr$), such that the measure,
referred to as the probability and denoted by $Pr$, of the sample space is 1. In Mhamdi's formalization of probability theory \cite{mhamdi_11}, given a probability space $p$, the functions \texttt{space}
and \texttt{subsets} return the corresponding
$\Omega$ and $\Sigma$, respectively. This formalization also includes the formal verification of some of the most widely used probability axioms, which play a pivotal role in formal reasoning about reliability properties.

Mathematically,
a random variable is a measurable function between a probability space and a
measurable space. A measurable space refers to a pair
($S,\mathcal{A}$), where $S$ denotes a set and $\mathcal{A}$ represents a nonempty collection of sub-sets of $S$. Now, if $S$ is a set with finite elements, then the corresponding random variable is termed as a discrete random variable and else it is called a continuous one. The
probability that a random variable $X$ is less than or equal to some value
$x$, $Pr(X \le x)$ is called the cumulative distribution function (CDF) and it characterizes the distribution of both discrete and continuous random variables. Mhamdi's formalization of probability theory \cite{mhamdi_11} also includes the formalization of random variables and the formal verification of some of their classical properties using the HOL theorem prover.

\section{Reliability}
\label{sec_3}
In reliability theory \cite{Bilinton_1992}, reliability $R(t)$ of a system or
component is defined as the probability that it performs its
intended function until some time $t$.
\begin{equation}
\label{reliability_eq} R(t) = Pr (X > t) = 1 - Pr (X \le t) = 1 - F_X(t)
\end{equation}
where $F_X(t)$ is the CDF. The random variable $X$, in the above definition, models
the time to failure of the system. Usually, this time to failure is
modeled by the exponential random variable with parameter $\lambda$
that represents the failure rate of the system. Now, the CDF can be modeled in HOL as follows:

\begin{flushleft}
\texttt{\bf{Definition 1: }} \label{CDF_def}
\emph{Cumulative Distributive Function} \\
\vspace{1pt} \texttt{$\vdash$ $\forall$  p X x. CDF p X x = distribution p X \{y | y $\leq$ Normal x\}
}
\end{flushleft}

\noindent where $p$ represents the probability space, $X$ is the random variable and $x$ represents a $real$ number. The function \texttt{Normal} converts a $real$ number to its corresponding value in the $extended-real$ data-type, i.e, the $real$ data-type including the positive and negative infinity. The function \texttt{distribution} accepts a probability space $p$, a random variable $X$ and a set and returns the probability of $X$ acquiring all the values of the given set in the probability space $p$. Now, Definition 1 can be used to formalize the reliability definition, given in Equation \ref{reliability_eq}, as follows:

\begin{flushleft}
\texttt{\bf{Definition 2: }} \label{Reliability_def}
\emph{Reliability} \\
\vspace{1pt} \texttt{$\vdash$ $\forall$  p X x. Reliability p X x = 1 - CDF p X x
}
\end{flushleft}

We used the above mentioned formal definition of reliability to formal verify some of the classical properties of reliability in HOL. The first property in this regard relates to the fact that the reliability of a good component is 1, i.e., maximum, prior to its operation, i.e., at time 0. This property has been verified in HOL as the following theorem.

\begin{flushleft}
\texttt{\bf{Theorem 1: }} \label{Reliability_AT_ZERO}
\emph{Maximum Reliability} \\
\vspace{1pt} \texttt{$\vdash$ $\forall$ p X. prob\_space p $\wedge$  (events p = POW (p\_space p)) $\wedge$ \\
\ \ \ \ \ \ \ ($\forall$ y. X y $\neq$ NegInf $\wedge$  X y $\neq$ PosInf) $\wedge$ \\
\ \ \ \ \ \ \ ($\forall$ z. 0 $\le$ z $\Rightarrow$ ($\lambda$x. CDF p X x) contl z) $\wedge$\\
\ \ \ \ \ \ \  ($\forall$ x. Normal 0 $\le$ X x) $\Rightarrow$\\
\ \ \ \ \ \ \ (Reliability p X 0 = 1)
}
\end{flushleft}

\noindent The first two assumptions of the above theorem ensure that the variable \emph{p} represents a valid probability space based on the formalization of Mhamdi's probability theory \cite{mhamdi_11}. The third assumption constraints the random variable to be well-defined, i.e., it cannot acquire negative or positive infinity values. The fourth assumption states that the CDF of the random variable $X$ is a continuous function, which means that $X$ is a continuous random variable. This assumption utilizes the HOL function \texttt{contl}, which accepts a lambda abstraction function and a real value and ensures that the function is continuous at the given value. The last assumption ensures that the random variable $X$ can acquire positive values only since in the case of reliability this random variable always models time, which cannot be negative. The conclusion of the theorem represents our desired property that reliability at \emph{time=0} is \emph{1}.

The proof of the Theorem 1 exploits some basic probability theory axioms and the following property according to which
the probability of a continous random variable at a point is zero.

The second main characteristic of the reliability function is its decreasing monotonicity, which is verified as the following theorem in HOL:

\begin{flushleft}
\texttt{\bf{Theorem 2: }} \label{Reliability_MONOTONE}
\emph{Reliability is a Monotone Function} \\
\vspace{1pt} \texttt{$\vdash$ $\forall$ p X a b.
prob\_space p $\wedge$  (events p = POW (p\_space p)) $\wedge$ \\
\ \ \ \ \ \ \ ($\forall$ y. X y $\neq$ NegInf $\wedge$  X y $\neq$ PosInf) $\wedge$ \\
\ \ \ \ \ \ \ ($\forall$ x. Normal 0 $\le$ X x) $\wedge$  a $\le$ b $\Rightarrow$\\
\ \ \ \ \ \ \  (Reliability p X (b)) $\leq$ (Reliability p X (a))
}
\end{flushleft}

\noindent The assumptions of this theorem are the same as the ones used for Theorem 1 except the last assumption, which describes the relationship between variables $a$ and $b$. The above property clearly indicates that the reliability cannot increase with the passage of time.

The formal reasoning about the proof of  Theorem 2 involves some basic axioms of probability theory and a property that the CDF is a monotonically increasing function.

Finally, we verified that the reliability tends to 0 as the time approaches infinity. This property is verified under the same assumptions that are used for Theorem 1.

\begin{flushleft}
\texttt{\bf{Theorem 3: }} \label{Reliability_TENDS}
\emph{Reliability Tends to Zero As Time Approaches Infinity} \\
\vspace{1pt} \texttt{$\vdash$ $\forall$ p X.  prob\_space p $\wedge$  (events p = POW (p\_space p)) $\wedge$ \\
\ \ \  ($\forall$ y. X y $\neq$ NegInf $\wedge$  X y $\neq$ PosInf) $\wedge$ ($\forall$ x. Normal 0 $\le$ X x) $\Rightarrow$\\
\ \ \ \ \ \ \  (lim ($\lambda$n. Reliability p X (\&n)) = 0)
}
\end{flushleft}

\noindent The HOL function \texttt{lim} models the limit of a real sequence. The proof of Theorem 3 primarily uses the fact that the CDF approches to 1 as its argument approaches infinity.

These three theorems completely characterize the behavior of the reliability function on the positive real axis as the argument of the reliability is time and thus cannot be negative. The formal verification of these properties based on our definition ensure its correctness. Moreover, these formally verified properties also facilitate formal reasoning about reliability of systems, as will be demonstrated in Section \ref{sec_5} of this paper. The proof details about these properties can be obtained from our proof script \cite{waqar_ftscs_13}.

\section{Formalization of Series Reliability Block Diagram}
\label{sec_4}

In a serially connected system \cite{Bilinton_1992}, depicted in Figure 1, the reliability of the complete system mainly depends upon the failure of a single component that has the minimum reliability among all the components of the system. In other words, the system stops functioning if any one of its component fails. Thus, the operation of such a system is termed as reliable at any time $t$, if all of its components are functioning reliably at this time $t$. If the event $A_{i}(t)$ represents the reliable functioning of the $i^{th}$ component of a serially connected system with $N$ components at time $t$ then the overall reliability of the system can be mathematically expressed as \cite{Bilinton_1992}:

\begin{equation}\label{eq2}
      R_{series}(t) = Pr (A_{1}(t) \cap A_{2}(t) \cap A_{3}(t) \cdots \cap A_{N}(t))
 \end{equation}

\begin{figure}\label{series_fig}
  \centering
  \includegraphics[width = 0.5\textwidth]{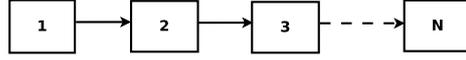}
  \begin{center}
  \caption{System with a Series Connection of Components}
  \end{center}\label{series_fig_caption}
\end{figure}

\noindent Using the assumption of mutual independence of individual reliability events of a series system \cite{Bilinton_1992}, the above equation can be simplified as:

\begin{equation}\label{eq3}
R_{series}(t) = \prod_{i=1}^{N}R_{i}(t)
\end{equation}

Moreover, an intrinsic property of a series system is that its overall reliability is always less than or equal to the reliability of the sub-component with the least reliability.

\begin{equation}\label{eq4}
R_{series}(t) \leq min(R_{i}(t))
\end{equation}

We proceed with the formalization of the series RBD by first formalizing the notion of mutual independence of more than two random variables, which is one of the most essential prerequisites for reasoning about the simplified expressions for RBD. Two events $A$ and $B$ are termed as mutually independent iff $Pr(A\cap B) = Pr(A)Pr(B)$. All the events involved in reliability modeling are generally assumed to be mutually independent. Since we often tackle the reliability assessment of systems with more than two components, we formalize the mutual independence of a list of random variables in this paper as follows:

\begin{flushleft}
\texttt{\bf{Definition 3: }} \label{mutual_indep_def}
\emph{Mutual Independence of Events } \\
\vspace{1pt} \texttt{$\vdash$ $\forall$ p L. mutual\_indep p L  = \\
\ \ $\forall$  L1 \vspace{1pt} n. PERM L L1 $\wedge$ 2 $\leq$ n $\wedge$ n $\leq$ LENGTH L $\Rightarrow$ \\
\ \ \ \ \ \ prob p (inter\_set p (TAKE n L1)) = \\ \ \ \ \ \ \ list\_prod (list\_prob p (TAKE n L1))
}
\end{flushleft}

\noindent The function \texttt{mutual\_indep} takes a list of events or sets $L$ along with the probability
space $p$ as input and returns True if the given list of events are mutually independent in $p$.
The formal definitions for the HOL functions used in the above definition are given in Table 1. The predicate \texttt{PERM} ensures that its two list arguments form a permutation of one another, the function \texttt{LENGTH} returns the length of a list, the function \texttt{TAKE} returns a list that contains the first $n$ elements of its argument list, the function \texttt{inter\_set} performs the intersection of all the sets in a list of sets and returns the probability space in case of an empty list argument, the function \texttt{list\_prob} returns a list of probabilities associated with the given list of events in the given probability space and the function \texttt{list\_prod} recursively multiplies all the elements of its argument list of real numbers. Thus, using these functions the function \texttt{mutual\_indep} ensures that for any 2 or more elements $n$, taken in any order, of the given list of events $L$, the property $Pr(\bigcap_{i=0}^nL_i) = \prod_{i=0}^nPr(L_i)$ holds.

\begin{table}
\begin{tabular}[c]{|l |l|}
\hline
\texttt{\textbf{Function Name}} \ \   & \texttt{\textbf{HOL Definition}}  \\ \hline
\hline
 \texttt{PERM}  &  \texttt{$\vdash$ $\forall$ L1 L2. PERM L1 L2 =} \\&  \ \ \ \texttt{$\forall$ x.
 FILTER (\$= x) L1 = FILTER (\$= x)L2 } \\ \hline
\texttt{LENGTH } &  \texttt{$\vdash$ (LENGTH [] = 0 )} $\wedge$ \\
		      &  \ \ \ \texttt{ $\forall$ h t. LENGTH (h::t) = SUC (LENGTH t)} \\
\hline
\texttt{TAKE } & \texttt{$\vdash$ ($\forall$ n. TAKE n [] = [])} $\wedge$ \\
               & \ \ \ \texttt{$\forall$ n x xs. TAKE n (x::xs) = if n = 0 then [] else } \\
               & \ \ \ \ \ \texttt{x::TAKE (n - 1) xs}\\
\hline
\texttt{inter\_set } & \texttt{$\vdash$ ($\forall$ p. inter\_set p [] =  p\_space p )} $\wedge$ \\
               & \ \ \ \texttt{$\forall$ p h t. inter\_set p (h::t) = h  $\cap$ inter\_set p t} \\
\hline
\texttt{list\_prod } & \texttt{$\vdash$ ($\forall$  list\_prod   [] =  1)} $\wedge$ \\
              & \ \ \ \texttt{$\forall$  h t. list\_prod   (h::t) = h  * list\_prod  t} \\
\hline
\texttt{list\_prob } & \texttt{$\vdash$ ($\forall$ p. list\_prob  p [] =  [])} $\wedge$ \\
               & \ \ \ \texttt{$\forall$ p h t. list\_prob  p (h::t) =} \\
               & \ \ \ \ \ \ \ \texttt{prob p (h $\cap$ p\_space p)  * list\_prob p t} \\
\hline
%\texttt{compl\_list } & \texttt{$\vdash$ ($\forall$ p. compl\_list  p [] =  [])} $\wedge$ \\
%               & \ \ \ \texttt{($\forall$ p h t. \texttt{ compl\_list p (h::t) =} \\
%               &  \ \ \ \ \ \ \ \texttt{p\_space p DIFF h $\cap$ p\_space p) :: compl\_list p t} \\

\texttt{min}  &  \texttt{$\vdash$ $\forall$ x y.  min x y = if x $\leq$ y then x else y }\\
\hline
\texttt{min\_rel } & \texttt{$\vdash$ ($\forall$ f. min\_rel  f [] =  1)} $\wedge$ \\
               & \ \ \ \texttt{$\forall$ f h t. min\_rel  f (h::t) = min (f h) (min\_rel f t)} \\
\hline

\end{tabular}\caption{HOL Functions used in Definition 3}
\end{table}

Next, we propose to formalize the RBDs in this paper by using a list of events, where each event models the proper functioning of a single component at a given time based on the corresponding random variable. This list of events can be modeled as follows:

\begin{flushleft}
\texttt{\bf{Definition 4: }} \label{list_RV_def}
\emph{Reliability Event List} \\
\vspace{1pt} \texttt{$\vdash$ $\forall$ p x. rel\_event\_list p [] x = [] $\wedge$ \\
\ \ $\forall$ p x h t. rel\_event\_list p (h::t) x = \\
 PREIMAGE h \{y | Normal x < y\} $\cap$ p\_space p :: rel\_event\_list p t x
}
\end{flushleft}

\noindent The function \texttt{rel\_event\_list} accepts a list of random variables, representing the time
to failure of individual components of the system, and a $real$ number $x$, which represents the time index
where the reliability is desired, and returns a list of sets corresponding to the events that the individual
components are functioning properly at the given time $x$. This list of events can be manipulated, based on the
structure of the complete system, to formalize various RBDs.

Similarly, the individual reliabilities of a list of random variables can be modeled as the following recursive
function:

\begin{flushleft}
\texttt{\bf{Definition 5: }} \label{list_reliability_function_def}
\emph{Reliability of a List of Random Variables } \\
\vspace{1pt} \texttt{$\vdash$ $\forall$ p x . rel\_list p [] x = [] $\wedge$\\
\ \ $\forall$ p h t x.  rel\_list p (h::t) x = \\ \ \ Reliability p h x :: rel\_list p t x
}
\end{flushleft}

\noindent The function \texttt{rel\_list} takes a list of random variables and a $real$ number $x$, which
represents the time index where the reliability is desired, and returns a list of the corresponding
reliabilities at the given time $x$. It is important to note that all the above mentioned definitions are
generic enough to represent the behavior of any RBD, like series, parallel, series-parallel
and parallel-series.

Now, using Equation (\ref{eq2}), the reliability of a serially connected structure can be defined as:

\begin{flushleft}
\texttt{\bf{Definition 6: }} \label{series_def}
\emph{System with a Series Connection of Components } \\
\vspace{1pt} \texttt{$\vdash$ $\forall$ p L.  rel\_series p L  =  prob p (inter\_set p L)
}
\end{flushleft}

\noindent The function \texttt{rel\_series} takes a list of random variables $L$, representing the failure times of the individual components of the system, and a probability space $p$ as input and returns
the intersection of all the events corresponding to the reliable functioning of these components using the
function \texttt{inter\_set}, given in Table 2. Based on this definition, we formally verified the result of Equation (\ref{eq2}) as follows:

\begin{flushleft}
\texttt{\bf{Theorem 4: }} \label{series_connected_system_def}
\emph{Reliability of a System with Series Connections} \\
\vspace{1pt} \texttt{$\vdash$ $\forall$ p L x. prob\_space p $\wedge$ (events p = POW (p\_space p)) $\wedge$\\
\ \ \ \ \  0 $\leq$ x $\wedge$
2 $\leq$ LENGTH (rel\_event\_list p L x) $\wedge$   \\
 \ \ \ mutual\_indep p (rel\_event\_list p L x) $\Rightarrow$ \\
 \  (rel\_series p (rel\_event\_list p L x) = list\_prod (rel\_list p L x))
}
\end{flushleft}

\noindent The first two assumptions ensure that $p$ is a valid probability space based on Mhamdi's probability
theory formalization \cite{mhamdi_11}. The next one ensures that the variable $x$, which models time,
is always greater than or equal to 0. The next two assumptions of the above theorem guarantee that we have a
list of at least two mutually exclusive random variables (or a system with two or more components).
The conclusion of the theorem represents Equation (\ref{eq2}) using Definitions 4 and 6. The proof of Theorem 4
involves various probability theory axioms, the mutual independence of events and the fact that the probability
of any event that is in the returned list from the function \texttt{rel\_event\_list} is equivalent to its
reliability. More proof details can be obtained from our proof script \cite{waqar_ftscs_13}.

Similarly, we verified Equation (4) as the following theorem in HOL:

\begin{flushleft}
\texttt{\bf{Theorem 5: }} \label{series_connected_system_min_reliability_def}
\emph{Reliability of a System depends upon the minimum reliability of the connected components} \\
\vspace{1pt} \texttt{$\vdash$ $\forall$ p L x.  prob\_space p $\wedge$ (events p = POW (p\_space p)) $\wedge$ \\
\ \ \ \ \ 0 $\leq$ x $\wedge$
2 $\leq$ LENGTH (rel\_event\_list p L x) $\wedge$   \\
 \ \ \ \ \ mutual\_indep p (rel\_event\_list p L x) $\Rightarrow$ \\
 \ \ \ \ \ \ (rel\_series p (rel\_event\_list p L x) $\leq$  \\
 \ \ \ \ \ \ \ \ min\_rel ($\lambda$ L. Reliability p L x) L)
}
\end{flushleft}

The proof of the Theorem 5 uses several probability theory axioms and the fact that any subset of a mutually independent set is also mutually independent.

The definitions, presented in this section, can be used to model parallel RBD \cite{Bilinton_1992} and formally verify the corresponding simplified reliability relationships as well. The major difference would be the replacement of the function \texttt{inter\_set} in Definition 6 by a function that returns the union of a given list of events.

\section{Reliability Analysis of a Pipeline System}
\label{sec_5}
A typical oil and gas pipeline can be partitioned into a series connection of $N$ segments,
where these segments may be classified based on their individual failure times.  For example, a 60
segment pipeline is analyzed in \cite{Zhang_08} under the assumption that the segments, which exhibit
exponentially distributed failure rates, can be sub-divided into 3 categories according to their failure
rates ($\lambda$), i.e., 30 segments with $\lambda= 0.0025$, 20 segments with $\lambda= 0.0023$ and
10 segments with $\lambda= 0.015$. The proposed approach for reliability analysis of pipelines allows us to
formally verify generic expressions involving any number of segments and arbitrary failure rates.
In this section, we formally verify the reliability of a simple pipeline, depicted in Figure 2, with $N$ segments having arbitrary exponentially distributed failure times.

\begin{figure}\label{series_pipeline_fig}
 \centering
  \includegraphics[width = 0.47\textwidth]{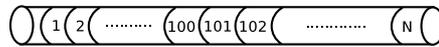}\\
  \caption{A Simple Pipeline}\label{series_pipeline_caption_fig}
\end{figure}

We proceed with the formal reliability analysis of the pipeline, shown in Figure 2, by formalizing the exponential random variable in HOL.

 \begin{flushleft}
\texttt{\bf{Definition 7: }} \label{Exponential_distribution_def}
\emph{Exponential Distribution Function } \\
\vspace{1pt} \texttt{$\vdash$ $\forall$ p X l. exp\_dist p X l = \\
\ \ \ $\forall$ x.  (CDF p X x = if 0 $\leq$ x then 1 - exp (-l * x) else 0)
}
\end{flushleft}

\noindent The predicate \texttt{exp\_dist} ensures that the random variable $X$ exhibits the CDF of an
exponential random variable in probability space $p$ with failure rate $l$. We classify a list of exponentially distributed random variables based on this definition as follows:

\begin{flushleft}
\texttt{\bf{Definition 8: }} \label{list_of exponential_distribution_function_def}
\emph{List of Exponential Distribution Functions} \\
\vspace{1pt} \texttt{$\vdash$ $\forall$ p L. list\_exp p [] L = T $\wedge$\\
\ \ $\forall$ p h t L.  list\_exp p (h::t) L = \\ \ \ exp\_dist p (HD L) h $\wedge$ list\_exp p t (TL L)
}
\end{flushleft}

\noindent The \texttt{list\_exp} function accepts a list of failure rates, a list of random variables $L$ and
a probability space $p$. It guarantees that all elements of the list $L$ are exponentially distributed with corresponding failure rates given in the other list within the probability space $p$. For this purpose, it utilizes the list functions \texttt{HD} and \texttt{TL}, which return the \emph{head} and \emph{tail} of a list, respectively.

Next, we model the pipeline, shown in Figure 2, as a series RBD as follows:

\begin{flushleft}
\texttt{\bf{Definition 9: }} \label{Reliab_series_def}
\emph{Reliability of Series Pipeline System } \\
\vspace{1pt} \texttt{$\vdash$ $\forall$ p L . pipeline p L = rel\_series p L
}
\end{flushleft}

\noindent Now, we can use Definition 8 to guarantee that the random variable list argument of the function \texttt{pipeline} contains exponential random variables only and thus verify the following simplified expression for the pipeline reliability.

\begin{flushleft}
\texttt{\bf{Theorem 6: }} \label{pipeline system}
\emph{Series Pipeline System  } \\
\vspace{1pt} \texttt{$\vdash$ $\forall$ p L x C. prob\_space p $\wedge$ (events p = POW (p\_space p)) $\wedge$ \\
\ \ \ \ \  0 $\leq$ x  $\wedge$  2 $\leq$ LENGTH (rel\_event\_list p L x) $\wedge$  \\
\ \ \ \ \    mutual\_indep p (rel\_event\_list p L x) $\wedge$  \\
\ \ \ \ \ \ list\_exp p C L $\wedge$  (LENGTH C = LENGTH L) $\Rightarrow$ \\
 \ \ \ \ \ \ (pipeline p (rel\_event\_list p L x) = exp (-list\_sum C * x))
}
\end{flushleft}

\noindent The first five assumptions are the same as the ones used in Theorem 5. The sixth assumption \texttt{list\_exp p C L} ensures that the list of random variable $L$ contains all exponential random variables with corresponding failure rates given in list $C$. The next assumptions guarantees that the lengths of the two lists $L$ and $C$ are the same. While the conclusion of Theorem 6 represents desired reliability relationship for the given pipeline model. Here the function \texttt{list\_sum} recursively adds the elements of its list argument and is used to add the failure rates of all exponentially distributed random variables, which are in turn used to model the individual segments of the series RBD of the pipeline. The proof of Theorem 6 is based on Theorem 4 and some properties of the exponential function \texttt{exp}. The reasoning was very straightforward (about 100 lines of HOL code)
compared to the reasoning for the verification of Theorem 4 \cite{waqar_ftscs_13}, which involved probability-theoretic
guidance. This fact illustrates the usefulness of our core formalization for conducting the reliability
analysis of pipelines.

The distinguishing features of this formally verified result include its generic nature, i.e., all the variables are universally quantified and thus can be specialized to obtain the reliability of the given pipeline for any given parameters, and its guaranteed correctness due to the involvement of a sound theorem prover in its verification, which ensures that all the required assumptions for the validity of the result are accompanying the theorem. Another point worth mentioning is that the individual failure rates of the pipeline segments can be easily provided to the above theorem in the form of a list, i.e., $C$. The above mentioned benefits are not shared by any other computer based reliability analysis approach for oil and gas pipelines and thus clearly indicate the usefulness of the proposed approach.

\section{Conclusions}
\label{sec_6}

Probabilistic analysis techniques have been widely utilized during the last two decades to assess the reliability of oil and gas pipelines. However, all of these probability theoretic approaches have been utilized using informal system analysis methods, like simulation or paper-and-pencil based analytical methods, and thus do not ensure accurate results. The precision of results is very important in the area of oil and gas pipeline condition assessment since even minor flaws in the analysis could result in the loss of human lives or heavy damages to the environment. In order to achieve this goal and overcome the inaccuracy limitation of the traditional probabilistic analysis techniques, we propose to build upon our proposed formalization of RBDs to formally reason about the reliability of oil and gas pipelines using higher-order-logic theorem proving.

Building upon the results presented in this paper, the formalization of other commonly used RBDs, including parallel, series-parallel and parallel-series, and the Weibull random variable is underway. These advanced concepts are widely used in the reliability analysis of pipelines. However, their formalization requires some advanced properties of probability theory. For example, for formalizing the reliability block diagrams of the series-parallel and parallel-series structures, we need to first formally verify the principle of inclusion exclusion \cite{Trivedi_02}. We also plan to formalize the underlying theories to reason about more realistic series pipeline systems, such as multi-state variable piping systems, where each subcomponent of the pipeline system consists of many irreversible states from good to worst. We also plan to investigate artificial neural networks in conjunction with theorem proving to develop a hybrid semi-automatic pipeline reliability analysis framework. Besides the pipeline reliability analysis, the formalized reliability theory foundation presented in this paper, may be used for the reliability analysis of a number of other applications, including hardware and software systems.

\section*{Acknowledgments}
This publication was made possible by NPRP grant \# [5 - 813 - 1 134] from the Qatar National Research Fund (a member of Qatar Foundation). The statements made herein are solely the responsibility of the author[s].
\bibliographystyle{splncs}
\bibliography{biblio}
\end{document}